\documentclass[aps,prl,reprint]{revtex4-1}

\usepackage{booktabs} 
\usepackage{array} 
\usepackage{paralist} 
\usepackage{verbatim} 
\usepackage{subfig} 
\usepackage{graphics}
\usepackage{graphicx}
\usepackage{lipsum}
\usepackage{graphicx}
\usepackage{color}

\newcommand{\includetxt}[1]{#1}
\newcommand{\excludetxt}[1]{}

\begin{document}


\title{Phonon-assisted robust and deterministic  two-photon biexciton preparation in a quantum dot}


\author{S. Bounouar\textsuperscript{1}, M. M\"{u}ller\textsuperscript{1}, A. M. Barth\textsuperscript{2}, M. Gl\"{a}ssl\textsuperscript{2}, V. M. Axt\textsuperscript{2} and P. Michler\textsuperscript{1}.}
\affiliation{\textsuperscript{1}Institut f\"{u}r Halbleiteroptik und Funktionelle Grenzfl\"{a}chen, Universit\"{a}t Stuttgart, Allmandring 3, 70569 Stuttgart,Germany.
\textsuperscript{2} Institut f\"{u}r Theoretische Physik III, Universit\"{a}t Bayreuth, Universit\"{a}tsstrasse 30, 95440 Bayreuth, Germany.}


\date{\today}

\begin{abstract}
\excludetxt{We investigate a more robust and  flexible alternative to Rabi oscillation-type biexciton preparation protocols traditionally used in semiconductor quantum dots. After a controlled shaping of strong laser pulses, the quantum dot is addressed from the biexciton two-photon resonance (Rabi oscillation regime)  to large positive energy detunings. Such a protocol makes use of the phonon-induced relaxation towards photon dressed states in optically driven quantum dots and combines the simplicity of traditional Rabi oscillation schemes with the robustness of adiabatic rapid passage schemes. It is shown that for excitation pulses in the picosecond range, one can find a detuning range where on demand initialization of the biexciton state is reached (fidelity $f_{XX}=0.98\pm 0.01$), together with a strong robustness to  pulse area fluctuations. Notably, the generated photons show similar coherence properties as measured in the resonant two-photon scheme. This protocol is a powerful and stable tool for the control of complex systems combining radiative cascades, entanglement, and resonant cavity modes in solid state systems.}

\includetxt{We investigate both experimentally and theoretically a simple yet more robust and flexible alternative to Rabi oscillation-type biexciton preparation protocols traditionally used for semiconductor quantum dots. The quantum dot is excited by a strong laser pulse positively detuned from the two-photon resonance yielding an on demand initialization of the biexciton state by making use of the phonon-induced thermalization of the photon dressed states.  It is shown that for excitation pulses in the picosecond range, a stable and high fidelity of up to $f_{XX}=0.98\pm 0.01$ is reached. Notably, the generated photons show similar coherence properties as measured in the resonant two-photon scheme. This protocol is a powerful tool for the control of complex solid state systems combining radiative cascades, entanglement and resonant cavity modes.}

\end{abstract}


\maketitle

\section{} One strong advantage of atomic-like systems, especially semiconductor
quantum dots (QD),  in the development of quantum computing or communication
devices is their ability to deliver on-demand single \cite{michler, He}  or
entangled photon pairs \cite{Muller,Akopian}. This first step towards the
possibility of deterministic quantum operations is a crucial complement to the
achievements obtained  in the field of quantum information processing
\cite{Bouwmeester, Nielsen} as well as for tests of fundamental aspects of
quantum mechanics  \cite{Haroche}. The initial state  is usually prepared
through a population inversion thanks to a strong coherent pulsed laser field
brought to resonance with the two-level system \cite{ Jayakumar, Brunner,
Stuffler}. This Rabi oscillation protocol can be very efficient, but  is strongly
sensitive to fluctuations of the excitation parameters like the pulse
area. Other more complex protocols, using chirped laser pulses to populate
adiabatically the upper state, are in principle more robust \cite{Wei},  but the
degree of population inversion realized in experiments   devoted to the
single exciton preparation \cite{Simon, Wu} stayed below the ideal case.  In
contrast to real atomic systems, solid state systems experience a coupling to
their environment, in particular with the surrounding crystal vibrational modes
\cite{Fedorov, Borri, besombes, Nazir, Kaer}. This has always been considered as a strong
limitation to their efficient use because of the occurring decoherence.  In
particular phonons have been identified as the cause for the non-ideal state
preparation using chirped pulse protocols \cite{Luker, Eastham, Debnath}.
However, it has been recently proposed for semiconductor quantum dots that, when
addressed with a controlled off-resonant pulse, this weakness can become an advantage
and make the state preparation more efficient, robust and  flexible \cite{Glassl}. 
The phonons cannot only be used to achieve an inversion in
simple two level systems consisting of
the ground  $|0\rangle$ and an exciton state $|X\rangle$ \cite{GlasslLongtime, Reiter, Hughes}, 
but also for the initialization of the biexciton state $|XX\rangle$ that forms
the upper level of a radiative cascade, which can potentially result in emission of entangled
photon pairs \cite{stevenson, Hafen}.

In this Letter, we investigate the biexciton preparation in a quantum dot
through a controlled  two-photon excitation scheme taking benefit of the
normally undesired carrier-phonon coupling. By tuning the laser energy and pulse
length, such that the relaxation processes due to this coupling are most
efficient, it is shown that one can transit from a resonant Rabi oscillation
regime to an adiabatic, efficient and robust biexciton state initialization. The
results for a sufficiently long pulse are compared to a situation where the
pulse length is too short to complete the relaxation and the consequences on the
state preparation are studied.  The coherence of the resulting emitted photons is
evaluated and reveals to be comparable in the Rabi $\pi$-pulse case and for the
phonon-assisted protocol, adding to efficiency and robustness, a long coherence
time. Accompanying our experiment, we have studied theoretically the dynamics of
a quantum dot driven by a laser with frequency close to the two-photon resonance
and we find a good overall agreement to the predictions of our simulations.
From this we can conclude that the underlying mechanism of the state preparation
is indeed the coupling to longitudinal acoustic phonons,  which we included in
our model to account for the solid state environment of the quantum dot.

\begin{figure*}
 \includegraphics[trim=2cm 4cm 3cm 4cm, clip=false, scale=0.55]{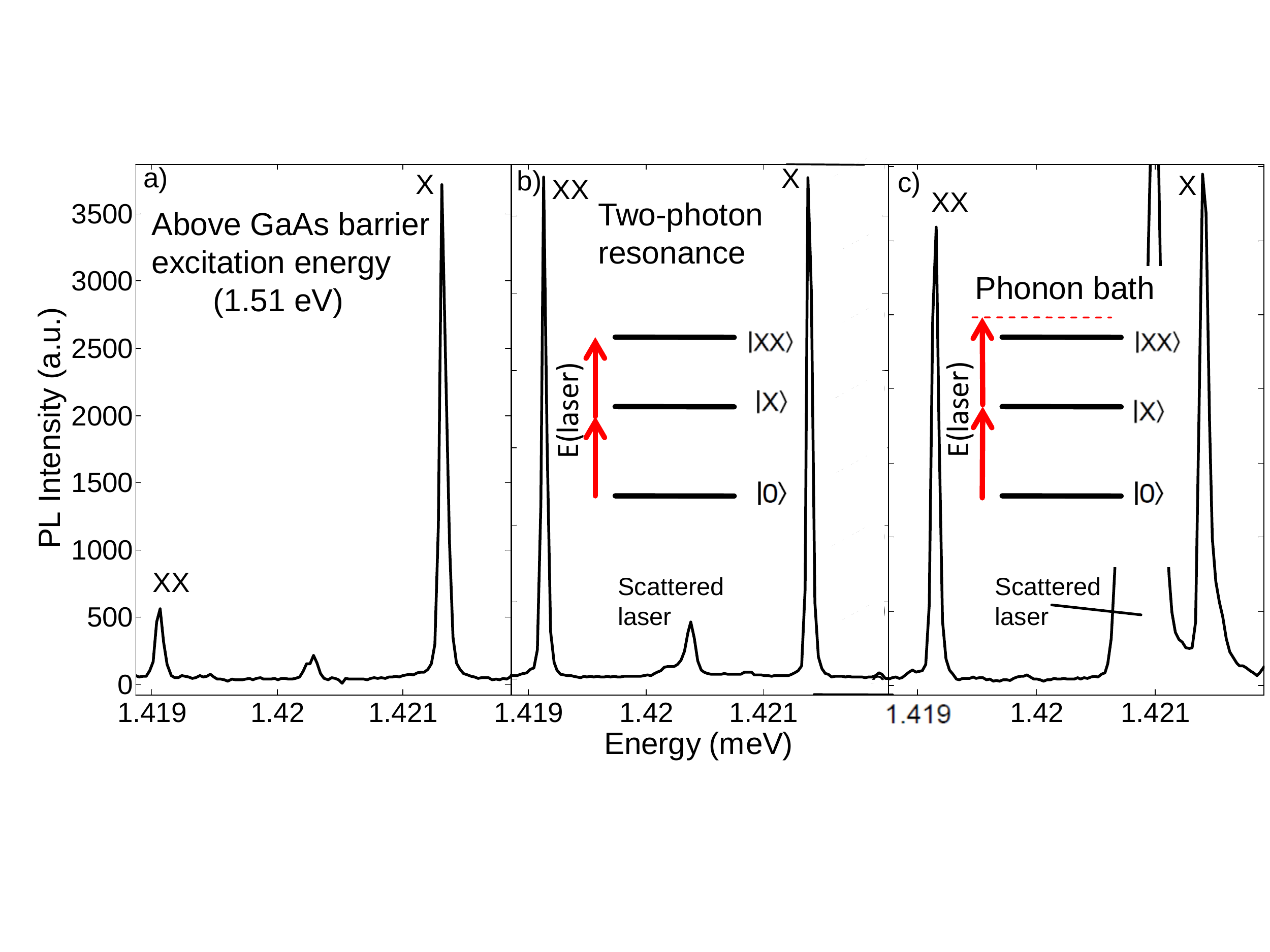}%
\captionsetup{justification=raggedright,
singlelinecheck=false
}
\caption{ Single QD emission spectrum a) under non resonant above bandgap excitation, b) under resonant two-photon biexciton state excitation, c) with phonon-assisted 13 ps pulsed, 0.65 meV positively detuned excitation. \label{fig:spectrumartikel}} 
 \end{figure*}
 
\begin{figure*}
 \includegraphics[trim=0cm 2cm 0cm 2cm, clip=true, scale=0.55]{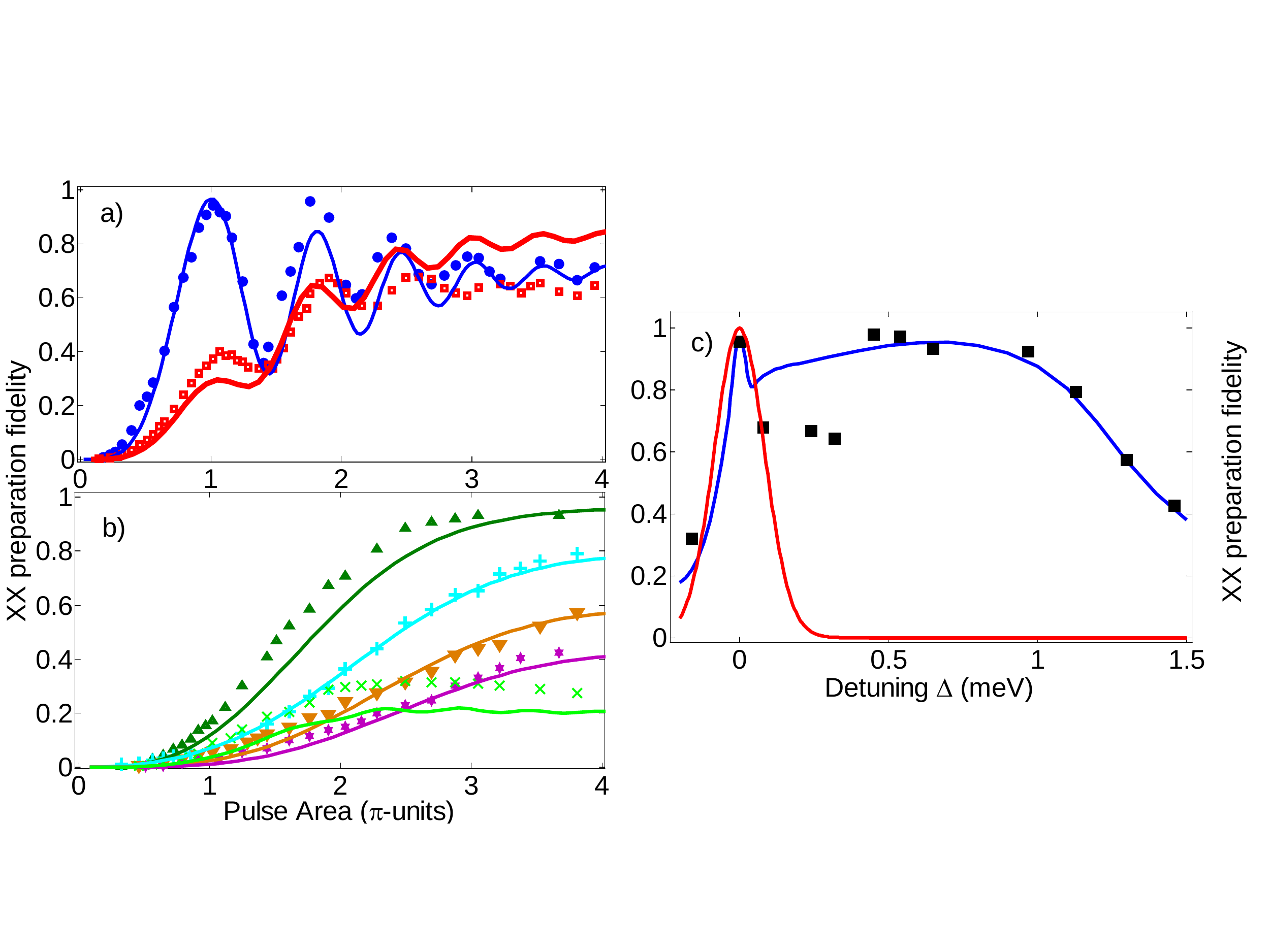}%

\captionsetup{justification=raggedright,
singlelinecheck=false
}
\caption{\small \sl QD  a) Biexciton occupation probability versus renormalized
pulse area with excitation pulse length of 13 ps superposed to path-integral
simulation results for different laser detunings from the TPBR: in resonance represented in
blue, 0.08 meV detuning in red, b) 0.65 meV detuning in green, 1.1 meV detuning
in light blue, 1.3 meV detuning in brown, 1.5 meV detuning in violet, -0.1 meV
detuning in light green, c) Maximum biexciton preparation fidelity (see text) reached for different laser detunings from the TPBR. The points represent the fidelities measured in the experiment
and the full lines are  results of the simulations. (in blue with phonon coupling included in the calculations, in red without any phonon coupling).
 \label{fig:powerdep13ps}} 
 \end{figure*}

We used an epitaxially grown (In,Ga)As/GaAs QD kept at a temperature of 4.2 K to
investigate the photoluminescence under pulsed two-photon excitation. In order
to address the biexciton in a resonant or a quasi-resonant way, by setting
the pulse length at a constant value and by controlling its wavelength, the laser is tailored through a pulse shaping setup. The
quantum dot is excited from the side with horizontally polarized light and the
detection is done after a perpendicularly oriented polarizer in order to reject
the scattered laser. The pulse length as well as the pulse shape was controlled
via intensity autocorrelation measurements. More details about the experimental
setup are given in the supplementary material. Figure \ref{fig:spectrumartikel}
a) shows a photoluminescence spectrum under above bandgap excitation. The
exciton (1.4212 eV) and biexciton lines (1.4189 eV) are separated by the
biexciton binding energy (2.3 meV). Direct excitation of the biexciton is
obtained by setting a shaped laser in resonance between the exciton and the
biexciton [see Fig.  \ref{fig:spectrumartikel} b)].  Although the biexciton and
the groundstate are not directly dipole coupled, the dynamics induced by a
resonant laser field results in Rabi-type oscillations where similar to a
directly coupled two-level system  the oscillation frequency scales with the
square-root of the laser intensity  \cite{Stuffler}.  Therefore, the final
inversion obtained after a pulse of finite length  also oscillates between
$\pm1$ as a function of the excitation power.  These Rabi oscillations were
probed with power dependent PL intensity measurements where the results for 13
ps pulses are shown as blue  dots  in Fig. \ref{fig:powerdep13ps} a).  The solid
lines are the result of a numerically exact real-time path-integral simulation
\cite{Vagov}, which gives us the ability to  solve our model of an optically
driven quantum dot  without further approximations taking into account arbitrary
multi-phonon processes as well as all non-Markovian effects. Because the size of
the quantum dot and the strength of the chirp of the laser pulse where not
directly measured, we treated these quantities as fitting parameters and matched
the scaling of the measured intensity to the first maximum of the theoretically
predicted Rabi oscillation for resonant excitation. Further information about
our model and other system parameters used in the calculations are given in the
supplementary material. 
 
The first maximum of the blue curve for resonant excitation in
Fig.~\ref{fig:powerdep13ps}~a) indicates the inversion of the biexciton
population and the corresponding pulse will be referred to as a $\pi$-pulse. At
the $\pi$-pulse power, the $|XX\rangle$ fidelity is estimated to be $f_{XX}=0.96\pm0.01$.
Two other noticeable features of these oscillations are the decrease of the Rabi
period  with increasing  pulse area and a damping of the amplitude.  The
first characteristic is a signature of the two-photon excitation process
\cite{Stuffler} and the damping is due to a coupling to  phonons
\cite{Forstner,Machnikowski}.  
The average value of the  oscillation is  above 0.5 due to an imperfection of the excitation pulse, 
which is weakly chirped  in the pulse shaping setup. For a positive sign of the chirp and low temperatures such
an increase of the biexciton population has already been predicted
\cite{GlasslChirp}. However, here it should be noted that according to our
simulations the frequency sweep mostly affects the dynamics for resonant
excitation, while for detuned laser pulses targeted in this Letter the chirp
only has a very small effect.

Let us now focus on the case of off-resonant excitation of the QD exemplarily
shown in the spectrum in Fig.~\ref{fig:spectrumartikel} c) for 0.65 meV
detuning. The laser detuning $\Delta$ is referred to the two-photon biexciton
resonance (TPBR) and in our measurements ranges from -0.1 meV up to 1.5 meV. For such
detuned excitation in a system with few isolated, discrete levels it is a well known result, that 
the amplitude as well as the mean value of the Rabi oscillations rapidly decreases
as the laser frequency is increased.  In fact, for a small positive detuning of
$\Delta=+0.08$ meV, shown in Fig.~\ref{fig:powerdep13ps}a) in red color, reduced
Rabi oscillations are still visible and when the detuning is increased they
disappear completely. However, as both our calculations and our experimental
data show, there is also an overall increase of the biexciton population for
positive detunings. This can be seen even more clearly in
Fig.~\ref{fig:powerdep13ps}~b), which shows the power dependence of the
biexciton population, for a few more selected detunings.  The state
initialization is  most efficient between 0.45 and 0.65 meV detuning (only the
0.65 meV detuning data are shown in dark green), where the measured $|XX\rangle$
preparation fidelity at high pulse area ($f_{XX}=0.98\pm 0.01$) is similar to
the one reached with the resonant $\pi$-pulse (0.96). In this range the fidelity
is not only stable against small changes of the excitation frequency, but also
shows a pronounced region where the biexciton population stays unaffected by
fluctuations of the laser power, which is a clear advantage compared to the
traditional resonant $\pi$-pulse scheme. For larger detunings (light blue for
1.1 meV, brown for 1.3 meV, violet for 1.5 meV), the preparation becomes less
efficient  consistent with our calculations. 

The detuning dependence of the biexciton fidelity can be seen in more detail in
Fig.~\ref{fig:powerdep13ps}~c), where we have plotted the maximum biexciton
population measured in the laser power range between 0$\pi$ and 4$\pi$ (black
dots) for the whole series together with the corresponding results of the
simulations (blue line). Also shown are results of calculations where the
exciton-phonon coupling was disregarded (red line). The strong influence of the
environment visible from the discrepancy between the blue and the red line on
the one hand, and the close overall agreement between the experimental data and
the calculations including the phonon interaction on the other hand, provide
clear evidence that the state preparation can be attributed to the
carrier-phonon coupling. To understand the physics behind this feature it is
important to note that a phonon-induced relaxation is possible as the bare
electronic states become dressed by the laser field. This relaxation can lead to
a thermal occupation of the photon dressed states, which for positive detunings
yields a high biexciton population \cite{Glassl}. 
For negative detunings (also shown in Fig.~\ref{fig:powerdep13ps}~b) in light green)
the energetic order of the dressed states changes, and thus the biexciton state is no
longer the final state of the relaxation at low temperatures, which explains
the steep decrease of the biexciton population seen
in Fig.~\ref{fig:powerdep13ps}~c) for $\Delta<0$. 
Around the two-photon resonance we can see a sharp peak in
Fig.~\ref{fig:powerdep13ps}~c) as the maximum fidelity in the observed pulse
area interval is determined by the height of the $\pi$-pulse peak.  When further
increasing the laser frequency the $|XX\rangle$ population drops down again until the
transition from the resonant Rabi oscillation scheme to the off-resonant
phonon-assisted state preparation occurs. In the experiments the population
stays below the calculated values for detunings between 0.08~meV and  0.32~meV,
but for higher detunings also reaches a wide plateau where the phonon-assisted
relaxation is most efficient and therefore ideal for deterministic state
initialization. The maximal efficiency of the phonon coupling in this region is
due to the resonance of the most pronounced phonon energies to the transitions
between the relevant dressed states. For even higher detunings the splitting
between the dressed states becomes too large and hence the phonon relaxation
does not take place efficiently yielding lower and lower values for the maximal
biexciton population as is nicely seen in both experiment and theory. It should
be noted that the resonance structure described above has the same origin as the
non-monotonic dependence of the phonon induced damping of Rabi oscillations on
the pulse area \cite{Machnikowski, VagovPRL, Ramsay}. 

\begin{figure}
 \includegraphics[trim=0.1cm 0.5cm 0cm 0.4cm, clip=true, scale=0.3]{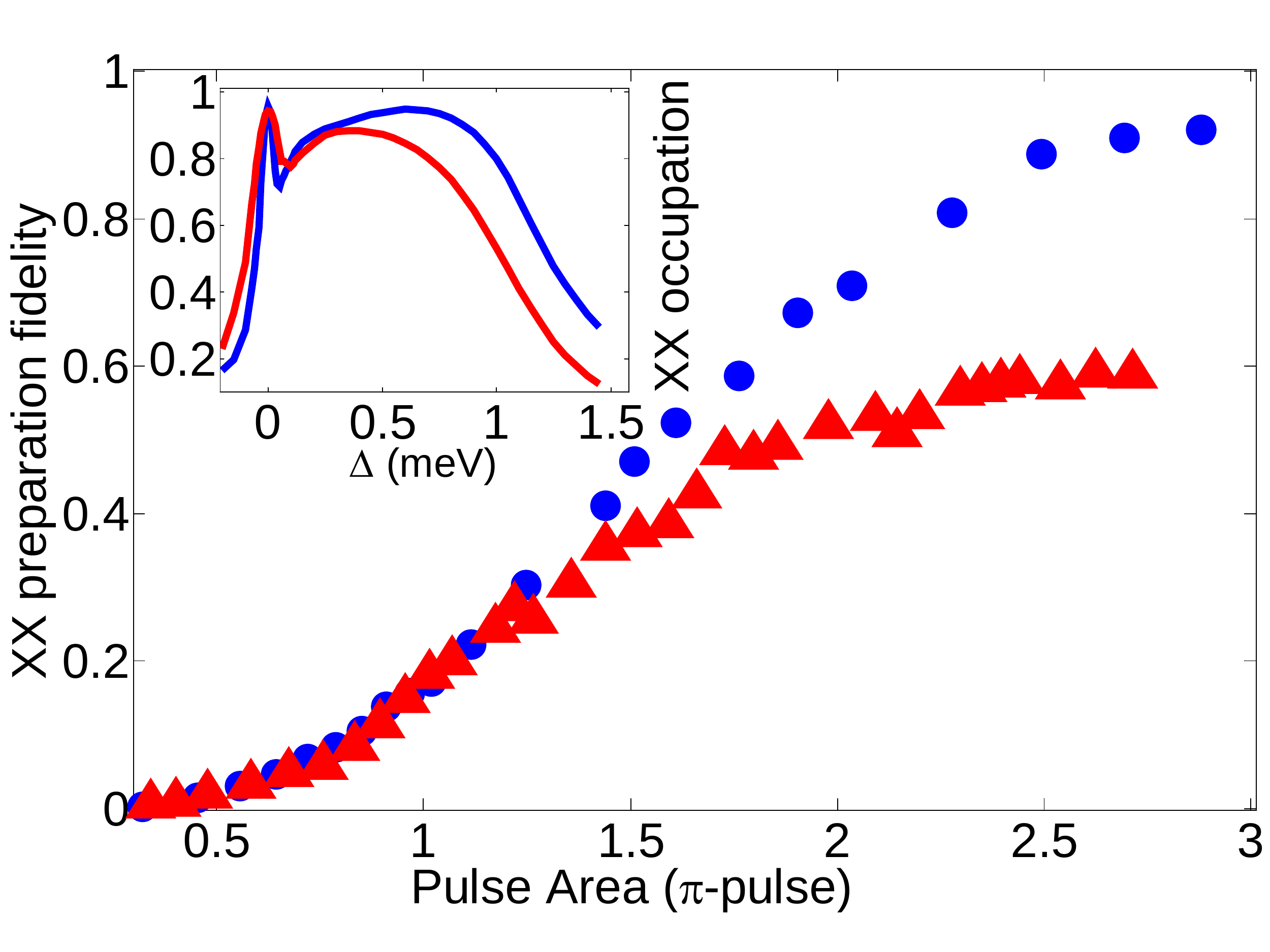}%

\captionsetup{justification=raggedright,
singlelinecheck=false
}

 \caption{\small \sl Measured biexciton population versus renormalized
excitation power for excitation pulse length of 13 ps (blue dots) and 7 ps (red
triangles) at laser detuning of 0.65 meV. Inset: Maximal biexciton occupation as
a function of the detuning for excitation with an unchirped pulse of length 13
ps (blue) and 7 ps (red). \label{fig:shorter_pulse}}
 \end{figure}

 In order to test the influence of the excitation pulse length, the same
experiment was carried out for shorter excitation pulses of 7 ps width.
Fig.~\ref{fig:shorter_pulse} shows the power dependencies at $\Delta=0.65$ meV
detuning for excitation pulse widths of 13 ps (blue dots)  and 7 ps (red
triangles), respectively. The biexciton preparation fidelity at high pulse areas becomes much less efficient for the shorter excitation pulse (7ps) and whatever the detuning used, the fidelity of the biexciton preparation obtained with short pulses stays insufficient (not shown).  The same tendencies
are also found in the theory as seen from the inset of
Fig.~\ref{fig:shorter_pulse}, where 
as  in Fig.~\ref{fig:powerdep13ps}~c), 
but for an unchirped excitation,
the detuning dependence of the 
maximal biexciton occupation  is plotted 
for pulse durations of
13 ps (blue) and 7 ps (red), respectively.  
While the maximal attainable
biexciton occupation is practically  independent of the pulse length for
resonant excitation, it is significantly reduced at larger positive detunings 
for the 7 ps pulse.
The calculations support the conclusion that robust and efficient preparation
can be obtained provided that the pulse is long enough to relax the system to
the energetically lowest dressed state during the pulse, whereas under too short
excitation no robust biexciton preparation can be achieved.

\begin{figure}
 \includegraphics[trim=0cm 2.5cm 0cm 2cm, clip=true, scale=0.35]{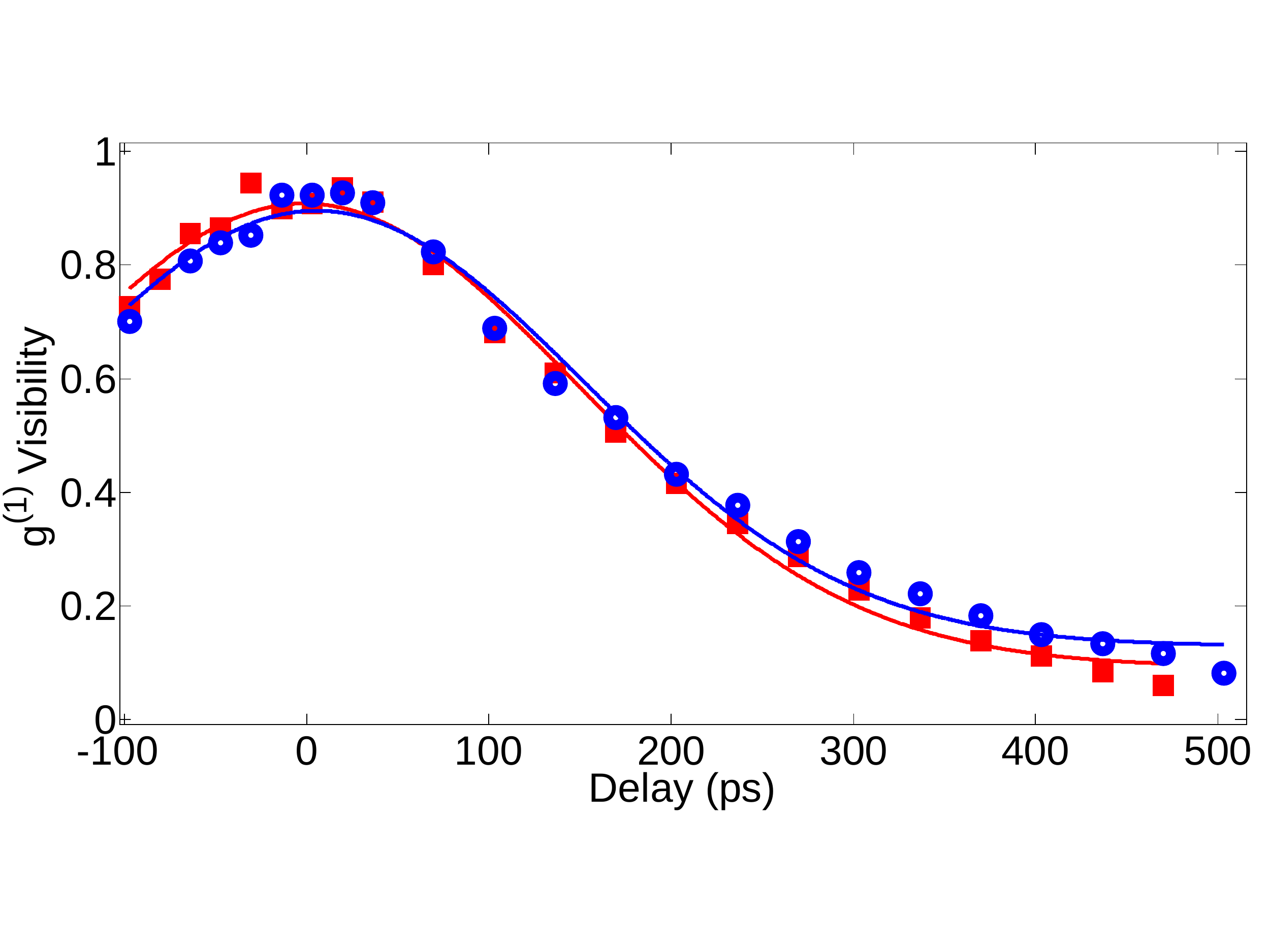}%

\captionsetup{justification=raggedright,
singlelinecheck=false
}
\caption{\small \sl Biexcitonic photon first order interference visibility
versus time delay, under resonant $\pi$-pulse excitation (blue dots), under
phonon-assisted, 0.65 meV detuned,  $3 \pi$-pulse excitation (red squares). Full
curves are Gaussian fits of the experimental data from which coherence times (
$267 \pm 5 $ ps in resonance, $260 \pm 6 $ ps for the 0.65 meV detuned excitation)
are extracted. \label{fig:coherence}} 
 \end{figure}

 In order to check the usefulness of the protocol in quantum optics
applications, we investigated the coherence of the generated photons and
compared them to coherence times observed in resonance. Fig. \ref{fig:coherence}
displays the measured coherence time  curves under 13 ps  resonant $\pi$-pulses
(blue)  and 0.65 meV detuned excitation (at large pulse area, i.e. around
$3\pi$, in red). They obviously show a very similar coherence time. The full
curves are Gaussian fits from which we extracted a coherence time of
$267 \pm 5$ ps and $260 \pm 6$ ps, respectively. This is in contrast with
measurements made with above bandgap excitation pulses which resulted in
significantly  lower coherence times ($\tau=114 \pm 4$ ps\cite{Muller}), because of the electronic fluctuations
generated in the surrounding of the quantum dots \cite{berthelot}. This
conservation of the coherence obtained in resonance is of crucial importance
since the coherence of the emitted photons is a decisive parameter
determining their degree of indistinguishability \cite{Santori, Trojani}.

In summary, it is demonstrated experimentally in this Letter that one can obtain a robust biexciton
preparation with near unity fidelity and a long coherence time by using a
simple protocol involving excitations detuned from the 
two-photon resonance. Comparing  with theoretical results we find a good agreement
 revealing that the preparation is  due to phonon-induced relaxation processes.
Applied to  the optical preparation of biexcitons in quantum dots this is a
particularly flexible and efficient scheme  for the initialization of entangled
photon states. 
Since this protocol leaves the TPBR free from laser scattering,
it is particularly suitable for a recently proposed two-photon
emission  ($|X\rangle$ and $|XX\rangle$) in a large Q-factor cavity mode set between $|X\rangle$ and $|XX\rangle$, which
was demonstrated as highly entangled whatever the $|X\rangle$ fine structure splitting is \cite{jahnke}. 
The approach presented in this Letter for the biexciton preparation 
would solve practical problems encountered in such
two-photon emission observations in QD-photonic crystals systems  \cite{Ota}
enabling clean photon statistic measurement.

Acknowledgment: 
The authors acknowledge L. Wang, A. Rastelli and O. Schmidt for
providing the high-quality sample. The authors acknowledge financial support
from the DFG via the projects MI 500/23-1 and AX 17/7-1.\\

\subsection{}
\subsubsection{}

\end{document}